\documentstyle[12pt]{article}
\author {Peter Huber \\Germanistisches Seminar\\
University of Heidelberg\\
Hauptstr. 207-209\\
D-69117 Heidelberg\\
Germany \\pethk@aol.com} 
\title {A Cosmologic Model Based on the Equivalence of
Expansion and Light Retardation \\ Part 2: Small-Scale
Aspects and Quanta} 
\begin{document}
\maketitle

\begin{abstract}
\noindent Interpretation of the cosmological red shift as
light retardation with the amount of $\dot c = -Hc$ yields
a photon rest mass $hH/c^2$. A system of natural units is 
introduced, in which the Planck mass is the
geometric average of the photon rest mass and the universal
mass. It is shown that the equivalence of expansion and
light retardation results in a Cosmologic Uncertainty
Principle (CUP), which determines the Heisenberg 
Principle. Within the Retarded Light Model (RLM) the
Dirac ``coincidences" are found to be systematic, and
an explanation is provided.
\end{abstract}

\section{Introduction}
The Retarded Light Model (RLM) as presented in (Huber
1992, Huber 1993 and Huber 2002) proceeds from the
fact that a cosmological decrease of the velocity of light
is possible within the Friedmann model. It has been found
a deceleration rate of
\begin{equation}
\dot c = - Hc \label{1}
\end{equation}
in order to explain the observed cosmological red shift
by light retardation. Since the Hubble parameter $H$ 
is associated with the relative expansion $\dot R/R$,
eq. (\ref{1}) can be written
\begin{equation}
{\dot R\over R} = H = -{\dot c\over c} \label{2}
\end{equation}
With a little rearrangement we find
\begin{equation}
\dot Rc + R\dot c = 0 \label{3}
\end{equation}
Integration yields the so-called Indiscernibility Principle
(IP)
\begin{equation}
Rc = const. \label{4}
\end{equation}
which states that the universe expands {\it only} in 
terms of a decreasing velocity of light. This occurs
according to the conveyor belt model: In the frame
of the light source the emitting frequency $\nu$
remains constant in
\begin{equation}
c=\lambda\nu \label{4a}
\end{equation}
while $\lambda$ decreases with $c$. However,
a light wave, once emitted, remains constant
while travelling through the universe and does
not participate in the cosmic expansion rate.
A remote receiver finds the frequency $\nu$
decreasing with $c$. Since light is subject to
gravitation it is suggested that
light retardation follows from the {\it eigen}
gravitation of the universe. While in General
Relativity, which deals with local gravitational
fields, the deceleration of light is transformed
into a temporal delay, this transformation does not
make sense when the universe as a whole is
considered, since it has been shown that the
emission frequencies of electromagnetic waves
are constant in the frame of the model. Thus a
defined emission frequency serves as a
standard clock of the Robertson-Walker-time
(the time which appears in the RWM). With
this preconditions the RLM yields a startling
consequence: If the expansion of the universe
is caused by light retardation and light retardation
is caused by gravitation, then the {\it expansion
is caused by gravitation}. This equivalence of
expansion and gravitation is referred to as
Generalized Equivalence Principle (GEP). With
the GEP it is obvious, why the observed expansion 
rate and the assumed proper gravitation of the 
universe seem to be in perfect equilibrium. 
Furthermore, the RLM is able to explain all the
classical problems of the standard model of
cosmology easily.

While part 1 of the series deals with purely
cosmological questions, this part is devoted to
small-scale and quantum aspects as implications
of the presented model. Hereby ``quantum aspects"
is meant in a wide sense; it mainly denotes the
appearence of the Planck constant $h$ which was
spared out in the first part.

\section{The Interaction Between the Photon and
the Universe}

In Special Relativity the photon has a rest mass of
zero. This is concluded from the fact, that any finite
value of mass will approach infinity while accelerating 
towards $c$. The situation in General Relativity should
be different, since light is subject to gravitation. Thus
the photon has a mass proportional to its frequency. But
where is the frame in Special Relativity at which a photon 
is ``at rest"? In the RLM there is no rest mass of the
photon but rather a ``minimal mass" $m_{ph}$, which 
follows from
\begin{equation}
m_{ph} = {h\nu\over c^2} \label{5}
\end{equation}
and the consideration that the lowest
possible circular frequency $\omega$ is one 
photon per universal age $1/H$. We then 
obtain with $\nu=\omega/2\pi=H/2\pi$ for
the photon mass a value of
\begin{equation}
m_{ph} = {\hbar H\over c^2} \label{6}
\end{equation}

Conceding a non-zero rest mass $m_{ph}$ to the photon 
it can be shown that the gravitational impact $F$ of the 
universe on the photon reproduces the correct amount of 
light retardation:
\begin {equation}
-F = -{m_{ph}MG\over R^2} = m_{ph} (-Hc) =
 m_{ph}\dot c  \label{7}
\end{equation}
This equation, where $M=c^3/GH$ is the universal 
mass (see part 1), $G$ the gravitational ``constant",
and $R=c/H$ the radius of the universe, reveals
again the identity of the gravitational interaction
between the universe and the photon on the one side
and the deceleration of the photon on the other side.

The Planck constant $h$ is independent of the 
cosmological ``constants" $c$, $G$ and $H$, which
vary with time. In the frame of the RLM it is set
constant. The cosmological term $H/c^2$, as it 
appears in eq. (\ref{6}), is also constant, because $c$ 
varies with $\sqrt{t}$ while $H$ varies with $t$ (see
part 1). That means, $m_{ph}$ does not increase its 
mass with cosmic expansion. With this suggested
mass definition a contradiction to Special Relativity 
can be avoided, since the constant $m_{ph}$ does 
not participate in mass increase due to acceleration like 
the regular mass, the university is composed of. (It
is just that fact of mass increase during acceleration
which forbids a non-zero photon mass in SR.)

There is another strong argument in ascribing a rest mass of
$hH/c^2$ to the photon. Because SR has $m_{ph} = 0$, the
Yukawa radius of light (i.e. its range) is infinite. In the
RLM we obtain a Yukawa radius $r_Y$ of
\begin{equation}
r_Y(m_{ph}) = {\hbar\over m_{ph}c} = {c\over H} = R 
\label{8}\end{equation}
This result, that light cannot reach further than the
``boundaries" of the universe, makes certainly more sense than
infinity: What happens, for example, when the extension of the
universe is slower than the speed of light? In fact, all the
horizon problems of the standard model result from this
question. As pointed out in part 1, the extension of the
 universal radius $R$ occurs in the RLM always with $c$.

\section{Metrics, Natural Scales, and the "Origin" of the
Universe}

Since the Retarded Light model is based on Friedmann's
equations, its metrics should be closely related to
Robertson-Walker-metrics (RWM). The original form of 
the RWM is
\begin{equation}
ds^2 = c^2 dt^2 - R(t)^2 f(r) \label{9}
\end{equation}
with $f(r) = dr^2 / (1-kr^2) + r^2(d\vartheta^2 + sin^2
\vartheta d\varphi^2)$. The only change we have to make is 
to replace $c^2$ by $c(t)^2$ in (\ref{45}). Then we get with
$c=c_0/\sqrt{1+2H_0t}$ and $R=R_0\sqrt{1+2H_0t}$ 
(see part 1):
\begin{equation}
ds'^2 = {c_0^2 dt^2\over 1+2Ht} - R_0^2 (1+2Ht) f(r) \label{10} 
\end{equation}
The two metrics deviate from each other for a large $t$. The
endeavours towards establishing metrics with the line element
$ds$ in the vicinity of our galaxy (cluster) will lead to systematic 
distortions at larger distances, as seen from $ds'$-metrics.

Beside $R$ there exist two more natural length scales in the 
universe. For the sake of distinction $R=c/H$ will be referred
to as $R_H$ from now on, where die Index $H$ can be
associated with the name of Hubble. Besides, there is the
Planck length $R_P$ which is given by
\begin{equation}
R_P = \sqrt{\hbar G\over c^3} \label{11}
\end{equation}
Finally we can define a length from the ``momentum"
of the universe $Mc = c^4/GH$ and the Planck constant
\begin{equation}
R_B = {\hbar\over Mc} = {\hbar GH\over c^4} \label{12}
\end{equation}
which can be according to the de Broglie relation interpreted
as the wave length of the universe. (The index $B$ stands
for the name of the French physicist.) These three 
natural units are related to each other:
\begin{equation}
R_H R_B = R_P^2 \label{13}
\end{equation}
The Planck length $R_P$ can be interpreted as the
geometric average of the Hubble radius $R_H$ and
the de Broglie wave-length of the universe $R_B$.
With a constant $\hbar$ and the variation laws for $c$,
$G$ and $H$ we obtain for the temporal
variation of $R_B$
\begin{equation}
\dot R_B = -HR_B \label{14}
\end{equation}
and
\begin{equation}
R_B(t) = {R_B(t_0)\over \sqrt{1+2H_0t}} \label{15}
\end{equation}
Differentiation of the Planck length $R_P$ gives
\begin{equation}
R_P(t) = const. \label{16}
\end{equation}
We find $R_H$ and $R_B$ varying inversely to
keep $R_P$ constant. The 
divergence of the introduced natural scales indicates
that atomic and cosmological processes evolve with
different ``speed".  This was already presumed by Dirac 
(1937, 1938, 1979).

Equation (\ref{13}) can be generalized as follows:
\begin{equation}
 X_H X_B = X_P^2, \hspace{1.5cm} (X\in \{R,T,M\}) 
\label{17} 
\end{equation}
where $R$ stands for radii, $T$ for time
and $M$ for masses. This covers all possible natural 
units. Their definitions are: 
\begin{equation} \begin{array}{ll}
\mbox{Hubble units:\ } & R_H=c/H \\
& T_H=1/H \\
&M_H=c^3/GH\\
\mbox{Planck units:\ } & R_P=(\hbar G/c^3)^{1/2} \\
& T_P=(\hbar G/c^5)^{1/2} \\ 
& M_P=(\hbar c/G)^{1/2}\\
\mbox{de Broglie units:\ } & R_B=\hbar GH/c^4 \\ 
& T_B=\hbar GH/c^5 \\
& M_B=\hbar H/c^2  \label{18}
\end{array}\end{equation}
where the Planck units are the geometric averages of the
others. The indices can be associated with numbers like
\begin{equation}
H:=1 \quad P:=0 \quad B:=-1\ , \quad (H,P,B)\ \in \ i
\label{19}
\end{equation}
Then we obtain the relation
\begin{equation}
\hbar = |\dot c| R_i T_i M_i \quad \mbox{for}\quad \sum i = 1 
\label{20}
\end{equation}
This means that the Planck constant $h$ can be defined by 
the absolute value of light retardation and a certain
combination of elementary units, for instance ($R_H, T_H,
M_B$) or ($R_P, T_P, M_H$). It looks like $h$ is rather a 
magnitude appearing with the general problem of measuring and 
not primarily a constant of the physics of elementary
particles.

Because of the variation of $R_H$ and $R_B$ there must exist
a time $\tau_{\alpha}$ where $R_H = R_B = R_P$.
Applying $R_H=R_0 \sqrt{1+2H_0t}$ and (\ref{15}) we
find
\begin{equation}
\tau_{\alpha} = {hGH\over 2c^5} - {1\over 2H} = 
{1\over 2}(T_B-T_H) \label{21}
\end{equation}
On the other hand, at time $\tau_{\alpha}$ there is also
$T_B=T_H$, so that, according to (\ref{21}), $\tau_{\alpha}
=0$ in the natural units. Thus we can regard the time 
$\tau_{\alpha}$, at which the natural units equalled each 
other, as the origin of the universe.

\section{Uncertainties}

In part 1, section 3 we have deduced a relation for 
the propagation of light in the RWM, which differs from 
the standard model. Equation (8) of part 1: $R(t_0)\nu_0
c(t_1) = R(t_1)\nu_1c(t_0)$ can be written
\begin{equation}
R(t)\nu(t) = c(t) \label{22}
\end{equation}
Replacing the photon frequency $\nu$ by the momentum
$p_{ph}=\hbar\nu/c$ we obtain
\begin{equation}
R_H p_{ph} = \hbar \label{23}
\end{equation}
This relation solved to $m_{ph}=\hbar /R_H c$ yields 
the photon mass $m_{ph}=M_B=\hbar H/c^2$ as 
defined above. We see that $m_{ph}$ follows directly 
from RWM in the interpretation of the RLM. There is 
no choice of assuming other values. It can be 
shown that this relation also holds for the relativistic
momentum $p(t) = mv/\sqrt{1-v^2/c^2}$ of free falling
massive particles. We replace the wave-length 
$\lambda_B$ by the frequency $\nu_B$ and obtain
the altered de Broglie relation 
\begin{equation}
\nu_B = mvc/\hbar  \label{24}
\end{equation}
and insert the expression for $\nu_B$ in eq. (\ref{22}).
We then obtain again with $mv=p$
\begin{equation}
R(t) p(t) = \hbar \label{25}
\end{equation}
where the momentum $p$ is now generalized for
photons and massive particles. The relation (\ref{25})
can be referred to as the Cosmological Uncertainty
Principle (CUP). What does it say? Let there be a
non-expanding rod at defined coordinates in an
expanding universe at time $t_0$. Where is the rod 
at time $t_1$, when the coordinates do not co-expand
with the universe? Obviously, the universe has
gained some more units of length. How to distribute
them? It is by no means a general solution to define 
the rod in the old and new system as sticking to the 
point (0,0,0), because any other object in relation to 
the rod must change its coordinates. Howsoever the
new coordinate system will be defined in homogeneous
space,  there is a uncertainty in location of any cosmic
object. And even more: because of the uncertain
location it is not at all clear, whether a change in
coordinates results by an arbitrary change of the
coordinate system or by a real physical movement. 
Therefore the velocity is also uncertain. So is, after 
SR, the mass of the object, since it depends on its 
velocity. So the CUP does not only say that the
momentum $p(t)$ decreases continuously with
expanding $R(t)$, it also says that the location
and the momentum of an object in the universe
can not be measured more accurately than $\hbar$.

There is yet another aspect to be mentioned. The
Indiscernibility Principle (IP) as introduced in 
section 3 of part 1 states that space expansion and
light retardation cannot be distinguished from
each other. Its quantitative expression is $Rc =
const.$ Multiplied with the photon mass $m_{ph}
= M_B$ we have
\begin{equation}
Rcm_{ph} = Rp_{ph} = \hbar \label{25a}
\end{equation}
This implies that IP and CUP are mathematically 
identical, which means in words that space 
expansion is identical with an uncertainty
in location while light retardation implies an
uncertainty in velocity and mass respectively in
momentum.

In the standard model there is no CUP, and even if it
would exist, there would not be a connection to
Heisenberg's Uncertainty Principle (UP). The reason
is once again that electromagnetic waves expand
in the empty cosmic space, however, not in matter 
conglomerations like galaxy clusters, which are tied 
together by gravitation. So the CUP would not have any 
effect on observations and measurements made on earth. 
In the RLM, on the other hand, there are no ``raisins in 
the pie". The cosmological light retardation takes place 
anywhere. (It has to be kept in mind, however, that, like 
in the standard model, the light delay of {\it local} 
gravitational fields is transformed into a temporal delay.) 
So the cosmic expansion takes place on earth just as well
as anywhere in empty space. The CUP must therefore 
be also effective on earth. This implies a connection
to the UP. Because of this cosmological uncertainty in
location and velocity, measurements on earth or
elsewhere cannot surpass the limits stated by the CUP.
Thus we will have in general Heisenberg's UP
\begin{equation}
\Delta x \Delta {\bf p}_x \ge \hbar \label{25b}
\end{equation}

In eq. (\ref{20}) we have found a relation for $\hbar$,
which is composed of the amount of light retardation
and a certain combination of natural units with the
restriction that the sum of indices must be 1.
Theoretically, there are possible combinations of 
natural units from $\sum i = -3$ to $\sum i = 3$. 
What can be said about the cases $\sum i \neq 1$? 
Introducing the quantity of the universal 
impact $h_u$, which shall have the definition 
\begin{equation}
h_u := E_H T_H = {c^5\over GH^2} \label{26}
\end{equation}
and which implies, analogously to the uncertainty relations, 
for any processes in the universe
\begin{equation}
\Delta E\Delta t \le h_u \label{27}
\end{equation}
then eq. (\ref{20}) yields for any indices $-3\le \sum i \le 
3$ the generalized relation
\begin{equation}
|Hc| R_i T_i M_i = \sqrt{h_u^{\Sigma i -1}\hbar^{3-\Sigma i}}
\label{28}
\end{equation}
These considerations imply that there may be universes
possible with other measurement restrictions than the UP.

\section{Dirac Numbers and Other ``Coincidences"} 

The quantities $G$, $\hbar$, $c$ and $H$ can be combined to a 
mass, which has the magnitude of a typical elementary
particle, such as the pion:
\begin{equation}
\left ( {\hbar^2H\over Gc}\right )^{1/3} \approx m_{\pi}
\label{29}
\end{equation}
(In the following we will use $m_{\pi}$ with the mean value 
of  eq.(\ref{29}) as a symbol for the average meson in particular 
and the average elementary particle in general.) This relation 
indicates that microphysical quantities may be influenced by 
the temporal state of the universe, which is represented by the 
Hubble parameter $H$. This consideration led to certain 
cosmological models like Dirac's {\it ansatz} from 1937 and 
1938, in which he proposed that relations like (\ref{29}) are 
fundamental though as yet unexplained truths. To keep the 
expression constant Dirac assumed a varying $G$ with 
$\dot G = -3GH$ (cf. also Weinberg 1972, p. 622). This 
is exactly the value the RLM obtaines from keeping the 
total energy of the universe constant (see part 1,
section 7). However, the numerical equality of 
$\dot G$ in Dirac's approach and in the suggested model
follows from independent considerations, since Dirac, for
instance,  assumed a constant $c$. Nevertheless, the RLM
is able to explain the existence of large numbers as will be
shown in the following.

In 1972 Weinberg confirmed, that eq. (\ref{29}) "relates a 
single cosmological parameter, $H_0$, to the fundamental
constants $\hbar$, $G$, $c$ and $m_{\pi}$, and is so far
unexplained" (p. 620). Using the natural masses defined in 
(\ref{18}) we can write (\ref{29}) as
\begin{equation}
\left ( {\hbar^2H\over Gc}\right )^{1/3} = (M_B\ M_P^2)^{1/3} 
= (M_B^2\ M_H)^{1/3}\label{30}
\end{equation}
It follows that it is possible to construct an infinite number of 
masses derived from the natural masses, such as $(M_B^2\
M_P)^{1/3}$, $(M_B\ M_P^3\ M_H^2)^{1/6}$, and so on. In
general:
\begin{equation}
m(i,j,k) = (M_B^i\ M_P^j\ M_H^k)^{(i+j+k)^{-1}} \quad , \quad 
i,j,k \in {\cal R}_0 \label{31}
\end{equation}
With respect to (\ref{17}) we can reduce (\ref{31}) to
\begin{equation}
m(i,j,k) = (M_B^{(i+j/2)}\ M_H^{(k+j/2)})^{(i+j+k)^{-1}} \quad , 
\quad i,j,k \in {\cal R}_0 \label{32}
\end{equation}

With (\ref{31}) we are able to approximate any 
particle mass if we use powers high enough. Eq. 
(\ref{30}) is characterized by low powers. That low
powers should be of high significance is obvious when we
consider $M_B$, $M_H$, $M_P=(M_BM_H)^{1/2}$, and
$m_{\pi}$. Let us,  for instance, construct another combination 
characterized by low powers, such as $(M_PM_H^2)^{1/3}$, 
which will be referred to as $M_{bl}$. With relation (\ref{29}) 
this mass can be transformed into
\begin{equation}
M_{bl} = (M_PM_H^2)^{1/3}= m_{\pi}\left ( {\hbar c\over 
Gm_{\pi}^2} \right )^{3/2} \label{33}
\end{equation}
There is an astonishing resemblance to the Chandrasekhar mass
\begin{equation}
M_c = m_n \left ( {\hbar c\over Gm_n^2} \right )^{3/2}
\label{34}
\end{equation}
where $m_n$ represents the neutron mass. In fact, $M_{bl}$
equals about 223 Chandrasekhar masses or 401 regular sun
masses. Black holes have more than 30 sun masses. Should
$M_{bl}$ represent the average mass of a black hole or even
the average mass of ``stars" in general, such as $m_{\pi}$
represents the average meson and particle mass? If so, then a
great deal of the dark matter would be concentrated in super
massive black holes. This would explain the lack of visible
matter in the Retarded Light universe with its density 
$2\varrho_c$, which ist twice the critical density of  the 
standard model (see part 1, section 9).

It has been often remarked, that the relation $\hbar c/Gm_n^2$
of the Chandrasekhar mass is in the order of $10^{40}$ and
represents the relation of strong force and gravitation. In fact,
its numerical value is $1.7\times 10^{38}$. We get an even
better value with the values appearing in $M_{bl}$:
\begin{equation}
{\hbar c\over Gm_{\pi}^2} = 3.8\times 10^{40} \label{35}
\end{equation}
with an assumed value for the Hubble parameter 
$H=2.5\times 10^{-18}s^{-1}$.

On the other hand we can describe particles as multiples of 
natural masses, such as the invariant quantity of $M_B$. 
Because of the constancy of $M_B$ and the temporal 
mass variation according to tle law $M_H(t)=M_H(t_0)(1+2H_0t)$ 
the quantities of masses expressed in units of $M_B$ will vary with 
the age of  the universe. In this sense we can assign another member 
to the $10^{40}$-family:
\begin{equation}
 m_{\pi} = 10^{40} M_B \label{36}
\end{equation}
which says that the quotient of the mean particle mass and the
photon mass is the Dirac number $\gamma = 10^{40}$. 

With (\ref{29}) we can write 
\begin{equation}
\gamma = m_{\pi}/M_B = (M_P^2/M_B^2)^{1/3} =
(M_H/M_B)^{1/3} \label{37}
\end{equation}
Because of
\begin{equation}
{d\over dt}\left ( {hH\over c^2}\right ) = \dot M_B = 0 
\label{38}
\end{equation}
the value of $\gamma$ depends on the temporal variation of 
$M_H$. It may be useful to ask at what time there was 
$\gamma = 1$, which implies the equality of electromagnetic 
and gravitational force, Hubble horizon  and classical electron 
radius as well as the equality of $m_{\pi}$ and $M_B$ on the 
one hand, $M_B$ and $M_H$ on the other, and much more. 
Setting $M_B = M_H = M_0(1+2H_0t)$ we find with 
$M_0=c_0^3/G_0H_0$ exactly the value of the turning 
point $\tau_{\alpha}$ as described in (\ref{21}). It turnes 
out that the value of $\gamma$ is a consequence of the 
universal state of evolution, just as Dirac assumed, and 
in this sense even an absolute, though variable, quantity. 
So it is not a coincidence that the average particle mass 
represented by $m_{\pi}$ can be expressed by $M_B$ 
and the Dirac number $\gamma$. 

According to the definitions of natural units (\ref{18}) and 
$h_u$ (\ref{26}) we have following relations:
\begin{equation}
{R_H\over R_B} = {M_H\over M_B} = {T_H\over T_B} =
{h_u\over \hbar} = {c^5\over\hbar GH^2} \approx 5.5\times
10^{121} \label{39}
\end{equation}
and
\begin{equation}
{R_H\over R_P} = {R_P\over R_B} = {M_H\over M_P} = 
{M_P\over M_B} = {T_H\over T_P} = {T_P\over T_B} = 
\sqrt{c^5\over\hbar GH^2} \approx 7.4\times 10^{60} 
\label{40}
\end{equation}
($H=2.5\times 10^{-18} s^{-1}$). These relations imply
following definition of $\gamma$:
\begin{equation}
\gamma:= \left ( {c^5\over\hbar GH^2}\right )^{1/3} =
3.8\times 10^{40}\label{41}
\end{equation}
With $M_H / M_B\approx 10^{121}$ and (\ref{36}) we 
obtain the slightly altered Dirac relation
\begin{equation}
{\makebox{universal mass}\over \makebox{average 
particle mass}} = 2\times 10^{81} \label{42}
\end{equation}
(Originally, Dirac used the proton mass and obtained the
dimensionless relation $10^{78}$.) Within the RLM we can 
give an exact deduction:
\begin{equation}
{M_H\over m_{\pi}} = {c^3\over GH}\left ({Gc\over
\hbar^2H}\right )^{1/3} = \left ({c^5\over\hbar
GH^2}\right)^{2/3} = \gamma^2 \label{43}
\end{equation} 
Accordingly we can deduce Dirac's distance relation $R_H/r_e 
\approx \gamma$, where $r_e$ is the radius of the electron. 
To do this we use the relations
\begin{equation}
m_e \approx \alpha m_{\pi} \label{44}
\end{equation}
where $\alpha$ is the fine structure constant, and
\begin{equation}
\alpha = {e^2\over 4\pi\varepsilon\hbar c} \label{45}
\end{equation}
We then obtain
\begin{equation}
{R_H\over r_e} \approx {c\over H}{4\pi\varepsilon c^2
m_{\pi}\alpha\over e^2} = \left ({c^5\over\hbar
GH^2}\right)^{1/3} = \gamma\label{46}
\end{equation}
There is no explanation hitherto why Dirac's mass relation 
is $\gamma^2$ while the distance relation is only $\gamma$. 
The RLM provides a good reason: Combining the Dirac
relation with the laws of mass and dirstance variation we
obtain
\begin{equation}
M_H = M_0(1+2H_0t) \approx 2\times 10^{81} m_{\pi}
 \label{47}
\end{equation}
and
\begin{equation}
R_H = R_0\sqrt{1+2H_0t} \approx 5.9\times 10^{40} r_e 
\label{48}
\end{equation}
We immediately see that the laws of temporal variation of $M$ 
and $R$ must yield $\gamma^2$ and $\gamma$ because of the
temporal variation according to $t$ and $\sqrt{t}$. 

Another quantity which is related to $\gamma$ is the fine 
structure constant of gravitation $\alpha_G$, as it appears in
the Chandrasekhar mass as a $10^{40}$-``coincidence". It 
plays the same role in the composition of stars as Sommerfeld's 
$\alpha$ in the composition of atoms. It is usually defined 
inversely as $10^{-40}$. Instead of $\alpha_G = m_n^2G/c\hbar$ 
we suggest the definition
\begin{equation}
\alpha_G:= {m_{\pi}^2G\over \hbar c} = \gamma^{-1} \label{49}
\end{equation}

It may be useful to ask if or how $\alpha$ and $\alpha_G$ vary 
with time. With the values for $c$, $G$ and $H$ as deduced in
section 4 of part 1 the temporal variation of $\gamma$ respectively
$\alpha_G^{-1}$ is
\begin{equation}
\dot\gamma = \dot\alpha_G^{-1} = {2\over 3}H\gamma \label{50} 
\end{equation}
and after integration
\begin{equation}
\gamma (t) = \gamma_0 (1+2H_0t)^{1/3} \label{51}
\end{equation}
Inserting $\tau_{\alpha} = 1/2 (T_B-T_H)$ we obtain exactly 
$\gamma=1$, which yields the important result that gravitational 
force and strong force were equal in strength at the origin of
the universe. Thus $\gamma$ is indeed a function of the 
universal age as Dirac assumed.

On the other hand we find for $\alpha=e^2/4\pi\varepsilon\hbar 
c$ with $e = const.$, $\hbar=const.$ and $\varepsilon$ varying 
inversely to $c$ (see section 9 of part 1, Huber 2002, and 
M{\o}ller 1972, p.416) the result
\begin{equation}
\alpha = const. \label{52}
\end{equation}
Although a constant $\alpha$ is expected by most
theoretical physicists, this result is a little surprising
because of the variation of $\alpha_G$ one could
have expected a variation of $\alpha$ as well. Now
we have to deal with the fact, that at a certain time
in the early universe the gravitational force must have
been stronger than the electromagnetical force. It is
not the place here to speculate how the universe must
have looked like then.

If the pion is multiplicated with $\alpha$ as in (\ref{44})
one roughly obtains the electron mass. If the pion is 
multiplicated with $\alpha_G$ one should accordingly 
obtain the graviton. Inserting the values we find
\begin{equation}
\alpha_G m_{\pi} = M_B \label{53}
\end{equation}
or the other way round
\begin{equation}
\gamma M_B = m_{\pi} \label{54}
\end{equation}
According to this the graviton and the de Broglie mass are 
equal and both identical with the photon rest mass. This had to
be expected, though, because both the electromagnetic and
the gravitational force spread through the whole universe.
Inserting $R_H$ for the Yukawa radius as in (\ref{8})
gives $M_B$ for the graviton mass as well as for the
photon rest mass. This result casts a new light on the 
unsuccessful approach of Einstein and others to 
regard elementary particles as spacetime singularities 
in which energy is trapped inside the gravitational radius 
of the particle. It turned out that the gravitational radius 
of the electron mass is smaller than the actual radius by 
about $10^{40}$. This enormous discrepancy was never 
made plausible. For that reason Weyl (1923) even considered 
Einstein's theory of gravitation as incomplete. As we have 
seen, towards the past or towards smaller regions of spacetime 
such as elementary particle size $\gamma$ converges 
against $1$. This can be shown in general with the
theoretical pion mass. Its gravitational radius $r_G$ is
\begin{equation}
r_G(m_{\pi}) = {Gm_{\pi}\over c^2} = {1\over\gamma}\left 
({\hbar G\over c^2H}\right )^{1/3} = \alpha_G 
(R_P^2 R_H)^{1/3} = {2 r_Y\alpha_G} \label{55}
\end{equation}

In case of $\gamma= \alpha_G = 1$ the gravitational 
radius is identical with the particle radius. As we have seen, 
the creation of elementary particles was only possible at a 
time when the gravitation force was of the same order as the 
strong force. Since that time only transformations between 
elementary particles are possible. This provides a good 
explanation why the baryon number is constant.

Especially the last two decades possible differences between 
atomic time and gravitational time have been intensively
discussed (Dirac 1978, Canuto and Goldman) again, and
scale-covariant theories of gravitation have been formulated 
(Canuto, Adams, et al.). We can express the difference of
gravitational time -- or as Dirac (1978) calls it -- "ephemeris 
time" $T_H = 1/H$ and atomic time
\begin{equation}
t_e = {e^2\over 4\pi\varepsilon m_e c^3} = {\alpha\hbar\over 
m_e c^2} \approx {\hbar\over m_{\pi}c^2} = (T_P^2 T_H)^{1/3}
\label{56}
\end{equation}
with the relation
\begin{equation}
{T_H\over t_e} = {m_e c^2\over \alpha\hbar H} = {m_e\over
\alpha M_B} \approx \gamma \label{57}
\end{equation}
Again we have used the numerical coincidence of
relation (\ref{44}).

\section{Summary}
The main result of this paper is the conclusion, that
Friedmann expansion identified with light retardation
yields Heisenberg's Uncertainty Relations. On this
base a unification of gravitational and quantum theory
could be possible. It was further shown that the
direction-independent {\it eigen} gravitation of the
universe causes a temporal delay of photons. It was
argued that the possible interpretations of this delay
(space expansion, light or time retardation) are
indiscernible and therefore equivalent. With the
introduction of natural units the numerical ``coincidences"
resulting from relations of micro- and macro-physical
quantities could be explained.

\end{document}